\documentclass{article}
\usepackage{amssymb}
\usepackage[centertags]{amsmath}

\begin{document}

\title{Transient behavior of the solutions to the second order difference equations by the renormalization method based on Newton-Maclaurin expansion}
\author{ Cheng-shi Liu \\Department of Mathematics\\Northeast Petroleum University\\Daqing 163318, China
\\Email: chengshiliu-68@126.com}

 \maketitle

\begin{abstract}

 The renormalization method based on  the Newton-Maclaurin expansion is applied to study the transient behavior of the solutions to the difference equations as they tend to the steady-states. The key and also natural step is to make the renormalization equations to be continuous such that the elementary functions can be used to describe the transient behavior of the solutions to difference equations. As the concrete examples, we deal with the important second order nonlinear difference equations with a small parameter. The result shows that the method is more natural than the multi-scale method.


\end{abstract}

\section{Introduction}

 Difference equations go into more and more important in modern science from theory and application. In practice, a large number of discrete models are derived out to describe the real physical phenomenon[1-3]. Therefore, the quantitative and qualitative studies on the difference equations become important and meaningful topics for researchers. Among those, the asymptotic behavior of the solutions is still an interesting subject from both of theory and practice. Many methods on the perturbation theory of difference equations have been proposed, and a lot of applications are given[4-17].

 In [18], I have given the renormalization method based on the Taylor expansion (TR, for simplicity) and corresponding homotopy renormalization method (HTR for simplicity), and applied them to obtain the global valid asymptotic solutions to a lot of the perturbation and non-perturbation problems.  the usual renormalization group (RG, for simplicity) method [19-21] and its geometrical formulations[22-25] can be derived from our renormalization method.  Further, in [26], we have generalized the renormalization  method to the difference equations. The proposed renormalization method is based on the Newton-Maclaurin expansion. We used the renormalization method to solve the steady-states solutions, and gave a lot of applications[26]. However, we do not consider the transient behavior of the solutions to difference equations. In a series of papers[27-30], Mickens studied the periodic solutions to the second order difference equations with a small parameter. In particular, in [30], Mickens introduced the multi-discrete time method to deal with the transient behavior of the solutions to difference equations, and obtained the limit cycles behaviors for some second order vibration problems. Mickens' method is beyond the Krylov-Bogoljubov method[30,31]. In Mickens' method, by introducing a new variable $s=\epsilon k$ where $k$ is the usual time discrete variable, one can obtain the differential equations of some quantities depending on $s$ to remove the secular terms in the perturbation series expansions. But the introduction of the variable $s$ is not very natural in some sense. In the paper, we use the renormalization method based on the Newtun-Maclaurin expansion to deal with the transient behavior of the solutions to difference equations, and obtain the corresponding limit cycles behavior for some second order vibration problems. Our method is more natural than the Mickens' multi-scale method since we need not to introduce the auxiliary parameter $s$.

This paper is organized as follows. In section 2, we summarize some main results on renormalization method based on the  Newton-Maclaurin expansion for asymptotic
analysis of difference equations.  In section 3, we use the proposed renormalization method
 to the transient behavior of the solutions for some vibration difference equations. The last section is a short conclusion.

\section{Outline of the renormalization method based on the Newton-Maclaurin expansion}

 We give an outline of the renormalization method for difference equations as follows. Given a sequence $y(n)$, its Newton-Maclaurin series can be given as
\begin{equation}
y(n)=y(m)+\binom{n-m}{1}\Delta y(m)+\binom{n-m}{2}\Delta^2 y(m)+\cdots+\binom{n-m}{r}\Delta^r y(m)+\cdots,
\end{equation}
where the difference operator is defined by
\begin{equation}
\Delta y(n)=y(n+1)-y(n),
\end{equation}
\begin{equation}
\Delta^r y(n)=\Delta(\Delta^{r-1}y(n)),  r=1,2,\cdots.
\end{equation}

For the following general linear or nonlinear difference equation
\begin{equation}
N(y(n))=\epsilon M(y(n)),
\end{equation}
where $N$ and $M$ are in general linear or nonlinear difference operators,
we suppose that the solution is expanded as a power series of the
small parameter $\epsilon$
\begin{equation}
y(n)=y_0(n)+y_1(n)\epsilon+\cdots+y_k(n)\epsilon^k+\cdots.
\end{equation}
Substituting it into the above equation gives
\begin{equation}
N(y_0(n))=0,
\end{equation}
and
\begin{equation}
N_1(y_1(n))=M_1(y_0(n)),\cdots,N_k(y_k(n))=M_k(y_{k-1}(n)),\cdots,
\end{equation}
and so on, for some operators $N_k$ and $M_k$. From these equations, we give
the general solution of $y_0$ and the
particular solutions of $y_k$, in which there are some
undetermined constants. Further, we expand these solutions as the  series at a general
point $m$ as follows,
\begin{equation}
y_j(n)=\sum_{k=0}^{+\infty}\Delta^{k}y_j(m) \binom{n-m}{k},j=0,1,2,\cdots.
\end{equation}
Then rearranging the summation of these series yields
the final solution
\begin{equation}
y(n,m)=\sum_{k=0}^{+\infty}Y_k(m,\epsilon) \binom{n-m}{k},
\end{equation}
where
\begin{equation}
Y_k(m,\epsilon)=\sum_{j=0}^{+\infty}\Delta^ky_{j}(m)\epsilon^j,
n=0,1,\cdots.
\end{equation}
From the formula (9),  we have

(i). $y(m)=y(m,m)=Y_0(m,\epsilon)$;

(ii).$Y_k(m,\epsilon)=\Delta^kY_{0}(m,\epsilon),k=1,2,\cdots$.

The formula (i) means that the solution is just given by the
first term $Y_0$ of the expansion when we consider $m$ as a
parameter, and hence all other terms need not be considered at all!
However, the first term includes some integral
constants to be determined.  The point is to take $r$
relations in case (ii) as the renormalization equations to determinate the $r$
unknown parameters. The whole theory can be summarized in the following three theorems and a definition( see[26] for details).

 \textbf{Theorem 1}. The exact solution of the
Eq.(4) is just
\begin{equation}
y(n)=Y_0(n,\epsilon),
\end{equation}
and furthermore, we have
\begin{equation}
\Delta Y_{k-1}(n,\epsilon)=Y_k(n,\epsilon),n=1,2,\cdots.
\end{equation}

\textbf{Theorem 2}. From the expansion (9), we have
\begin{equation}
\Delta_my(n,m)=0,
\end{equation}
where $\Delta_m$ is the partial difference operator, that is, $\Delta_my(n,m)=y(n,m+1)-y(n,m)$.

\textbf{Theorem 3}. Consider vector difference equation
\begin{equation}
\Delta Y(n)=F(Y(n), \epsilon),
 \end{equation}
where $Y$ is a vector and $F$ is a vector value function.  We assume that $\widetilde{Y}(n,m)$ is the local solution at the general point $m$ and satisfies
 \begin{equation}
\Delta \widetilde{Y}(n,m)=F(\widetilde{Y}(n,m), \epsilon)+O(\epsilon^k).
 \end{equation}
Then we have
 \begin{equation}
\Delta \widetilde{Y}(m,m)=F(\widetilde{Y}(m,m), \epsilon)+O(\epsilon^k).
 \end{equation}

\textbf{Definition 1}. We call the relations
\begin{equation}
Y_k(m,\epsilon)=\Delta^kY_{0}(m,\epsilon),k=1,2,\cdots
\end{equation}
as the renormalization equations.

The detailed proofs and applications can be found in [26].

\section{The transient behavior of the solutions to second order nonlinear difference equations}
 Consider the second order nonlinear difference equation
 \begin{equation}
 z(n+1)-2z(n)+z(n-1)=(\Delta t)^2(-z(n)-\epsilon f(z(n+1),z(n),z(n-1))),
 \end{equation}
where $\epsilon$ is a positive small parameter. Further, write it as
\begin{equation}
 z(n+1)-(2-(\Delta t)^2)z(n)+z(n-1)=(\Delta t)^2\epsilon f(z(n+1),z(n),z(n-1))).
 \end{equation}
Assume that the solution can be expanded as a power series of the
small parameter $\epsilon$
\begin{equation}
z(n)=z_0(n)+z_1(n)\epsilon+\cdots+z_k(n)\epsilon^k+\cdots.
\end{equation}
Substituting it into the above equation yields the equations of
$z_k(n)$'s such as
\begin{equation}
 z_0(n+1)-(2-(\Delta t)^2)z_0(n)+z_0(n-1)=0,
\end{equation}
and
\begin{equation}
 z_1(n+1)-(2-(\Delta t)^2)z_1(n)+z_1(n-1)=(\Delta t)^2f(z_0(n+1),z_0(n),z_0(n-1)),
\end{equation}
and so forth. Solving the first equation gives
\begin{equation}
z_0(n)=A(1+i\Delta t)^n+B(1-i\Delta t)^n.
\end{equation}
Then the second equation becomes
\begin{equation}
z_1(n+1)-(2-(\Delta t)^2)z_1(n)+z_1(n-1)=(\Delta t)^2f(A(1+i\Delta t)^n+B(1-i\Delta t)^n).
\end{equation}
Here and after, for simplicity, we write $f(z_0(n+1),z_0(n),z_0(n-1))=f(z_0(n))$. If $f$ is an analytic function, we can expand it as a power series
 \begin{equation*}
f(A(1+i\Delta t)^n+B(1-i\Delta t)^n)=f(0)+f'(0)(A(1+i\Delta t)^n+B(1-i\Delta t)^n)
\end{equation*}
 \begin{equation*}
+\frac{f''(0)}{2}(A^2(1+i\Delta t)^{2n}+B^2(1-i\Delta t)^{2n}+2AB(1+i\Delta t)^n(1-i\Delta t)^n)+\cdots
\end{equation*}
 \begin{equation*}
=f(0)+f'(0)(A(1+i\Delta t)^n+B(1-i\Delta t)^n)+\frac{f''(0)}{2}\{A^2(1+i\Delta t)^{2n}
\end{equation*}
 \begin{equation}
+B^2(1-i\Delta t)^{2n}+2AB\}+\cdots
\end{equation}
Then by the variation of constant method, we  will obtain the particular solution of $z_1$. Here we choose two concrete examples to compute it.

\textbf{Example 1}. Take $f(z)=-z^3$. Then we have
\begin{equation*}
f(z_0(n))=-A^3(1+i\Delta t)^{3n}-B^3(1-i\Delta t)^{3n}-3A^2B(1+i\Delta t)^{2n}(1-i\Delta t)^n
\end{equation*}
\begin{equation}
-3AB^2(1-i\Delta t)^{2n}(1+i\Delta t)^n.
\end{equation}
By the variation of constant method, we have the particular solution of $z_1$,
\begin{equation*}
z_1(n)=-(\Delta t)^2\{\frac{A^3(1+i\Delta t)^{3n}}{(1+i\Delta t)^{3}+(1+i\Delta t)^{-3}-2+(\Delta t)^2}
\end{equation*}
\begin{equation*}
+\frac{B^3(1-i\Delta t)^{3n}}{(1-i\Delta t)^{3}+(1-i\Delta t)^{-3}-2+(\Delta t)^2}
\end{equation*}
\begin{equation}
+\frac{3A^2Bn(1+i\Delta t)^{n}}{(1+i\Delta t)-(1+i\Delta t)^{-1}}+\frac{3AB^2n(1-i\Delta t)^{n}}{(1-i\Delta t)-(1-i\Delta t)^{-1}}\}.
\end{equation}
Therefore, the solution is given by
\begin{equation*}
z(n)=A(1+i\Delta t)^n+B(1-i\Delta t)^n
\end{equation*}
\begin{equation*}
-\epsilon(\Delta t)^2\{\frac{A^3(1+i\Delta t)^{3n}}{(1+i\Delta t)^{3}+(1+i\Delta t)^{-3}-2+(\Delta t)^2}
\end{equation*}
\begin{equation*}
+\frac{B^3(1-i\Delta t)^{3n}}{(1-i\Delta t)^{3}+(1-i\Delta t)^{-3}-2+(\Delta t)^2}
\end{equation*}
\begin{equation}
+\frac{3A^2Bn(1+i\Delta t)^{n}}{(1+i\Delta t)-(1+i\Delta t)^{-1}}+\frac{3AB^2n(1-i\Delta t)^{n}}{(1-i\Delta t)-(1-i\Delta t)^{-1}}\}+O(\epsilon^2).
\end{equation}
in which the last two terms are secular terms.

By the renormalization method, considering two constants $A$ and $B$  as the functions of variable $m$,  the renormalization equations can be taken as
\begin{equation}
\Delta A(m)=\frac{3}{2}\epsilon\mathrm{i}\Delta t A^2(m)B(m),
\end{equation}
\begin{equation}
\Delta B(m)=-\frac{3}{2}\epsilon\mathrm{i}\Delta t B^2(m)A(m).
\end{equation}
From these two equations, we have $\Delta (AB)=0$ and hence $AB=c$ where $c$ is a constant. Hence the renormalization equations become
\begin{equation}
\Delta A(m)=\frac{3}{2}\epsilon\mathrm{i}c\Delta t A(m),
\end{equation}
\begin{equation}
\Delta B(m)=-\frac{3}{2}\epsilon\mathrm{i}c\Delta t B(m).
\end{equation}
Solving these two equations give
\begin{equation}
A(m)=A_0(1+\frac{3}{2}\epsilon\mathrm{i}\Delta t)^m,
\end{equation}
\begin{equation}
B(m)=B_0(1-\frac{3}{2}\epsilon\mathrm{i}\Delta t )^m.
\end{equation}
Then, we get the global asymptotic solution
\begin{equation*}
z(n)=A_0(1+\frac{3}{2}\epsilon\mathrm{i}\Delta t)^n(1+i\Delta n\Delta t)^n+B_0(1-\frac{3}{2}\epsilon\mathrm{i}\Delta t)^m(1-i\Delta t)^n
\end{equation*}
\begin{equation}
-\frac{\epsilon}{5}(A^3_0\exp\{\frac{9}{2}\epsilon c\mathrm{i}n\Delta t)(1+i\Delta t)^{3n}+B^3_0\exp(-\frac{9}{2}\epsilon c\mathrm{i}n\Delta t)(1-i\Delta t)^{3n}\}+O(\epsilon^2).
\end{equation}
Now  we  can calculate the transient behavior of the solutions as $n\rightarrow+\infty$ with keeping $n\Delta t=t$. In fact we have
\begin{equation*}
z(t)=A_0\exp((\frac{3}{2}\epsilon c+1)\mathrm{i} t)+B_0\exp(-(\frac{3}{2}\epsilon c+1)\mathrm{i}t)
\end{equation*}
\begin{equation}
-\frac{\epsilon}{5}(A^3_0\exp\{3(\frac{3}{2}\epsilon c+1)\mathrm{i} t)+B^3_0\exp(-3(\frac{3}{2}\epsilon c+1)\mathrm{i} t))+O(\epsilon^2),
\end{equation}
where $B_0=\overline{A_0}$ and $c=|A_0|^2$. This is a periodic solution which describes the limit cycles behavior.

Now we take another trick to solve the renormalization equations. Indeed, we denote $t=n\Delta t$, $A(m)=A(\frac{m}{n}\Delta t)$ and $B(m)=B(\frac{m}{n}\Delta t)$, and  take the limit $\Delta t \rightarrow 0$, we obtain the differential equations
 \begin{equation}
A'(t)=\frac{3}{2}\epsilon\mathrm{i} A^2(t)B(t),
\end{equation}
\begin{equation}
B'(t)=-\frac{3}{2}\epsilon\mathrm{i}A(t) B^2(t),
\end{equation}
From the two equations, we have
\begin{equation}
AB=c,
\end{equation}
where $c$ is a constant. And then, we solve it and give
\begin{equation}
A(t)=A_0\exp(\frac{3}{2}\epsilon c\mathrm{i}t),
\end{equation}
\begin{equation}
B(t)=B_0\exp(-\frac{3}{2}\epsilon c\mathrm{i}t),
\end{equation}
where $A_0$ and $B_{0}$ are  two arbitrary constants. From these solutions, we have
\begin{equation}
A(n)=A_0\exp(\frac{3}{2}\epsilon c\mathrm{i}n\Delta t),
\end{equation}
\begin{equation}
B(t)=B_0\exp(-\frac{3}{2}\epsilon c\mathrm{i}n\Delta t).
\end{equation}
Then, we give the global asymptotic solution
\begin{equation*}
z(n)=A_0\exp(\frac{3}{2}\epsilon c\mathrm{i}n\Delta t)(1+i\Delta n\Delta t)^n+B_0\exp(-\frac{3}{2}\epsilon c\mathrm{i}n\Delta t)(1-i\Delta t)^n
\end{equation*}
\begin{equation}
-\frac{\epsilon}{5}(A^3_0\exp\{\frac{9}{2}\epsilon c\mathrm{i}n\Delta t)(1+i\Delta t)^{3n}+B^3_0\exp(-\frac{9}{2}\epsilon c\mathrm{i}n\Delta t)(1-i\Delta t)^{3n}\}+O(\epsilon^2).
\end{equation}
Now  we  take $n\rightarrow+\infty$ with keeping $n\Delta t=t$ and obtain the same asymptotic solution (36).

\textbf{Example 2}. We take $f(z(n))=z(n+1)-z(n-1)$, that is, we consider a Van der Pol type of perturbation difference equation
 \begin{equation}
 z(n+1)-(2-(\Delta t)^2) z(n)+z(n-1)=\epsilon\Delta t(1-z^2(n))(z(n+1)-z(n-1)),
 \end{equation}
where $\epsilon$ is a positive small parameter. This equation is a center finite difference approximation of the
continuous Van der Pol differential equation[4,30]. Assume that the solution can be expanded as a power series of the
small parameter $\epsilon$
\begin{equation}
z(n)=z_0(n)+z_1(n)\epsilon+\cdots+z_k(n)\epsilon^k+\cdots.
\end{equation}
Substituting it into the above equation yields the equations of
$y_k(n)$'s such as
\begin{equation}
z_0(n+1)-(2-(\Delta t)^2) z_0(n)+z_0(n-1)=0,
\end{equation}
and
\begin{equation}
 z_1(n+1)-(2-(\Delta t)^2) z_1(n)+z_1(n-1)=\Delta t(1-z_0^2(n))(z_0(n+1)-z_0(n-1)),
\end{equation}
and so forth. Solving the first equation gives
\begin{equation}
z_0(n)=A(1+i\Delta t)^n+B(1-i\Delta t)^n.
\end{equation}
By the variation of constant method, we solve the second equation and obtain
\begin{equation*}
z_1(n)=\Delta t\{\frac{A^3\{(1+i\Delta t)^{3n-1}-(1+i\Delta t)^{3n+1}\}}{(1+i\Delta t)^{3}+(1+i\Delta t)^{-3}-2+(\Delta t)^2}
\end{equation*}
\begin{equation*}
+\frac{B^3\{(1-i\Delta t)^{3n-1}-(1-i\Delta t)^{3n+1}\}}{(1-i\Delta t)^{3}+(1-i\Delta t)^{-3}-2+(\Delta t)^2}
\end{equation*}
\begin{equation*}
+\frac{(A-A^2B)n\{(1+i\Delta t)^{n+1}-(1+i\Delta t)^{n-1}\}}{(1+i\Delta t)-(1+i\Delta t)^{-1}}
\end{equation*}
\begin{equation}
+\frac{(B-AB^2)n\{(1-i\Delta t)^{n+1}-(1-i\Delta t)^{n-1}\}}{(1-i\Delta t)-(1-i\Delta t)^{-1}}\}
\end{equation}
\begin{equation*}
=\frac{2}{5}A^3i(1+i\Delta t)^{3n}-\frac{2}{5}B^3i(1-i\Delta t)^{3n}
\end{equation*}
\begin{equation}
+\Delta t(A-A^2B)n(1+i\Delta t)^{n}-\Delta t(B-AB^2)n(1-i\Delta t)^{n}.
\end{equation}
Therefore, the solution is given by
\begin{equation*}
z(n)=A(1+i\Delta t)^n+B(1-i\Delta t)^n
\end{equation*}
\begin{equation*}
+\epsilon\{\frac{2}{5}A^3i(1+i\Delta t)^{3n}-\frac{2}{5}B^3i(1-i\Delta t)^{3n}
\end{equation*}
\begin{equation}
+\Delta t(A-A^2B)n(1+i\Delta t)^{n}-\Delta t(B-AB^2)n(1-i\Delta t)^{n}\}.
\end{equation}
in which the last two terms are secular terms.

By the renormalization method, considering two constants $A$ and $B$  as the functions of variable $m$,  the renormalization equations can be taken as
\begin{equation}
\Delta A(m)=\epsilon\Delta t (A(m)-A^2(m)B(m)),
\end{equation}
\begin{equation}
\Delta B(m)=\epsilon\Delta(B(m)- B^2(m)A(m)).
\end{equation}
In order to get the real solutions, we must take $B=\overline{A}$ in which case the above two equations are compatible. Further, we take $A=A_1+A_2i$ to obtain
\begin{equation}
\Delta A_1(m)=\epsilon\Delta t A_1(m)(1-A_1^2(m)-A_2^2(m)),
\end{equation}
\begin{equation}
\Delta A_2(m)=\epsilon\Delta t A_2(m)(1-A_1^2(m)-A_2^2(m)),
\end{equation}
from which we have
\begin{equation}
A_2=cA_1,
\end{equation}
where $c$ is a constant.

Now we take a key step, that is, we denote $t=n\Delta t$, $A(m)=A(\frac{m}{n}\Delta t)$ and $B(m)=B(\frac{m}{n}\Delta t)$, and  take the limit $\Delta t \rightarrow 0$, we obtain the differential equations
 \begin{equation}
A_1'(t)=\epsilon A_1(t)(1-(1+c)A_1^2(t)).
\end{equation}
From the equation, we have
\begin{equation}
A_1(t)=\frac{A_0e^{\epsilon t}}{\sqrt{1+(1+c)A_0^2e^{2\epsilon t}}},
\end{equation}
\begin{equation}
A_2(t)=\frac{A_0ce^{\epsilon t}}{\sqrt{1+(1+c)A_0^2e^{2\epsilon t}}}
\end{equation}
where $A_0$ is an arbitrary constant. Then, we give the global asymptotic solution
\begin{equation*}
z(n)=\frac{A_0e^{\epsilon t}}{\sqrt{1+(1+c)A_0^2e^{2\epsilon t}}}\{(1+ic)(1+i\Delta t)^n+(1-ci)(1-i\Delta t)^n\}
\end{equation*}
\begin{equation}
+\epsilon\frac{2}{5}i\frac{A^3_0e^{3\epsilon t}}{(1+(1+c)A_0^2e^{2\epsilon t})^{3/2}}\{(1+ic)^3(1+i\Delta t)^{3n}+(1-ci)^3(1-i\Delta t)^{3n}\}
\end{equation}
Now  we  can calculate the transient behavior of the solutions as $n\rightarrow+\infty$ with keeping $n\Delta t=t$. In fact, we have the asymptotic solution
\begin{equation*}
z(t)=\frac{A_0e^{\epsilon t}}{\sqrt{1+(1+c)A_0^2e^{2\epsilon t}}}\{(1+ic)e^{it}+(1-ci)e^{-it}\}
\end{equation*}
\begin{equation}
+\epsilon\frac{2}{5}i\frac{A^3_0e^{3\epsilon t}}{(1+(1+c)A_0^2e^{3it})^{3/2}}\{(1+ic)^3e^{3it}-(1-ic)^3e^{-3it}\}.
\end{equation}
 This is a periodic solution which describes the limit cycles behavior.

\textbf{Remark 1}: In Mickens' papers[27-30], what he studied is the following form difference equations
\begin{equation}
\frac{x_{k+1}-2x_k+x_{k-1}}{4\sin^2(\frac{h}{2})}+x_k=\epsilon f(x_{k+1},x_k,x_{k-1}.
\end{equation}
When $h$ tends to zero, we know that $4\sin^2(\frac{h}{2})$ can be replaced by $h^2$. Further, by taking $h=\Delta t$, we just give the same equation (18). Eq.(18) is just the discretization of the differential equation
\begin{equation}
z''(t)+z(t)=\epsilon f(z(t),z'(t)).
\end{equation}

\textbf{Remark 2}. In example 2, we should take $f(z(n))=\frac{z(n+1)-z(n-1)}{2}$, and then $\epsilon$ is replaced by
$\frac{\epsilon}{2}$. This means that in continuous model we take $f(z)=z'$ to describe the nonlinear vibration with small friction action.

\section{Conclusions}

From the applications we can see that in some examples, a key step is to transform the discrete  difference equations into the  differential equations by which we can easily solve the solutions and then derive the transient behavior of the solutions. Indeed, the reason of nonlinear difference equations being more difficult than nonlinear differential equations is just because of lacking effective representation of the solutions in terms of elementary functions. For example, for the simple nonlinear difference equations such as $\Delta y(n)=y(n)(1-y^2(n))$, we cannot give its explicit solution by elementary functions. When the considered difference equation is derived from the differential equation by discretization, we can naturally use the renormalization method to study it. From our examples, it is easy to see that the proposed method is valid for such difference equations. On the other hand, if the considered difference equations  is not from the discretization of some differential equation, or the corresponding remormalization equations are nonlinear difference equations which cannot be reduced to some simple solvable forms,  how to calculate its transient behavior  of the solution need still to consider in future.

\textbf{Acknowledgments}. I would like to thank Dr.Cameron Hall(University of Limerick) for his asking me how to use my renormalization method to deal with  the transient behavior of the solutions to difference equation so that I begin to think about the problem and write the paper.


\begin{thebibliography}{2}

\bibitem{14}R E Mickens. Difference Equations. Reinhold, New York,
1987.

\bibitem{1}R P Agarwal. Difference Equations and Inequalities: Theory,
Methods, and Applications. Dekker, New York, 1992.
\bibitem{5}H H  Holmes. Introduction to Perturbation Methods.
Springer, New York, 1995.

\bibitem{18}A E Nayfeh. Perturbation Methods. Wiley¨CInterscience,
New York, 1973.
 \bibitem{16}J A  Murdock. Perturbations, Theory and Methods. Wiley,
New York, 1991.
\bibitem{17}R E Jr O'Malley. Singular Perturbation Methods for Ordinary
Differential Equations. Springer, New York, 1991.

\bibitem{3}I V Andrianov,L I Manevitch. Asymptotology: Ideas,
Methods, and Applications. Kluwer Academic, Dordrecht,
2002.
\bibitem{22}F Verhulst. Methods and Applications of Singular Perturbations.
Springer, New York, 2005.
\bibitem{23}L Jodar,J L Morera. Singular Perturbations for Systems of Difference Equations.
 Appl. Math. Lett. 1990, 3:51-54.
\bibitem{11}A Luongo. Perturbation methods for nonlinear autonomous
discrete-time dynamical systems. Nonlinear
Dyn. 1996, 10:317-331.
\bibitem{12}A Maccari. A perturbation method for nonlinear two dimensional
maps. Nonlinear Dyn. 1999, 19:295-312.
\bibitem{19}T Sari, T Zerizer. Perturbations for linear difference
equations. J. Math. Anal. Appl. 2005, 305:43-52.
\bibitem{13}A Marathe, A Chatterjee. Wave attenuation in nonlinear
periodic structures using harmonic balance and multiple
scales. J. Sound Vib. 2006, 289: 871-888.
\bibitem{CTP522}W T Van Horssen, M C Ter Brake. On the multiple scales perturbation method for difference
equations. Nonlinear Dyn. 2003, 55:401-418.
\bibitem{CTP522}M Rafei, W T Van Horssen. Solving systems of nonlinear difference equations
by the multiple scales perturbation method. Nonlinear Dyn. 2012, 69:1509-1516.
\bibitem{rr}T Kunihiro, J  Matsukidaira. Dynamical reduction of discrete systems based on the renormalization-group method. Physical Review E, 1998, 57(4): 4817.
\bibitem{cam}C L Hall, C J Lustri. Multiple scales and matched asymptotic expansions for the discrete logistic equation. Nonlinear Dynamics, 2016, 85(2): 1345-1362.

\bibitem{r}Cheng-shi Liu. The renormalization method based on the Taylor expansion and applications for asymptotic analysis. Nonlinear Dyn. 2017, 88:1099-1124.


\bibitem{l4}N Goldenfeld, O, Martin, Y Oono. Intermediate asymptotics and renormalization group theory.
 J. Sci. Comput.  1989, 4:355-372.

\bibitem{g}L Y Chen, N Goldenfeld, Y Oono. Renormalization group and singular perturbations: Multiple scales,
 boundary layers, and reductive perturbation theory. Phys. Rev. E.  1996, 54:
376-394.

\bibitem{k}L Y Chen, N Goldenfeld, Y Oono. Renormalization group theory for global asymptotic analysis.
  Phys. Rev. Lett. 1994, 73: 1311-1315.


\bibitem{l5}T Kunihiro. A geometrical formulation of the renormalization group method for global analysis.
 Prog. Theor. Phys.  1995, 94:503-514.
\bibitem{bu3}T Kunihiro. A geometrical formulation of the renormalization group method for global analysis
II: Partial differential equations.
 Japan J. Indus. Appl. Math.  1997, 14: 51-69.
\bibitem{bu}T Kunihiro. The renormalization-group method applied to asymptotic analysis of vector fields.
Prog. Theor. Phys. 1997, 97: 179-200.





\bibitem{k}T Kunihiro. Renormalization-group resummation of a divergent series of the perturbative wave functions of the quantum anharmonic oscillator. Physical Review D, 1998, 57(4): R2035.
\bibitem{l2}Cheng-shi Liu. The renormalization method for asymptotic analysis: from differential equations to difference equations. arXiv preprint arXiv:1702.08512v2, 2017.

\bibitem{m4}R E Mickens. Periodic solutions of second-order nonlinear difference equations containing a small parameter. Journal of the Franklin Institute, 1985, 316(3): 273-277.
\bibitem{m3}R E Mickens. Periodic solutions of second-order nonlinear difference equations containing a small parameter¡ªII. Equivalent linearization. Journal of the Franklin Institute, 1985, 320(3-4): 169-174.

\bibitem{m2}R E Mickens. Periodic solutions of second-order nonlinear difference equations containing a small parameter¡ªIII. Perturbation theory. Journal of the Franklin Institute, 1986, 321(1): 39-47.
\bibitem{15}R E Mickens. Periodic solutions of second order nonlinear
difference equations containing a small parameter-IV.
Multi-discrete time method. J. Franklin Inst. 1987, 324: 263-271.





\bibitem{7}R L Huston. Krylov-Bogoljubov method for difference
equations. SIAM J. Appl. Math. 1979, 19:334-339.




\end{thebibliography}
\end{document}